\documentclass[12pt,a4paper]{article}
\usepackage{amsmath}
\usepackage{amssymb} 
\usepackage{graphicx}
\begin{document}
\begin{titlepage} 
\begin{flushright} IFUP--TH--2015\\ 
\end{flushright} ~
\vskip .8truecm 
\begin{center} 
\Large \bf Polyakov relation for the sphere and higher genus surfaces 
\end{center}
\vskip 1.2truecm 
\begin{center}
{Pietro Menotti} \\
\vskip 0.5truecm  
{\small\it Dipartimento di Fisica, Universit{\`a} di Pisa}\\ 
{\small\it e-mail: pietro.menotti@unipi.it}\\ 
\end{center} 
\vskip 1.2truecm
                
\vskip 1.2truecm

\begin{abstract}

The Polyakov relation, which in the sphere topology gives the changes
of the Liouville action under the variation of the position of the
sources, in the case of higher genus is related also to the
dependence of the action on the moduli of the surface. We write and
prove such a relation for genus 1 and for all hyperelliptic
surfaces.

\end{abstract}

\end{titlepage}

\eject

\section{Introduction}

On the sphere topology the Polyakov relation connects the dependence
of the action on the position of the sources with the accessory
parameters of the related Riemann-Hilbert problem. Such a relation was
originally conjectured by  Polyakov exploiting the semiclassical limit
of quantum operator product expansion \cite{polyakov}.

The relation plays a key role in several fields related to Liouville
theory like the hamiltonian formulation of $2+1$-dimensional gravity
\cite{CMS1,CMS2,CMS3} and the study the conformal block expansion of
the quantum correlation functions \cite{ZZ,piatek,hadasz1,hadasz2}.
The accessory parameters appear in the generalized monodromy problem
\cite{LLNZ,NRS} also in connection with the Nekrasov-Shatashvili limit
of super Yang-Mills theory \cite{piatek,ferraripiatek} and the AGT
conjecture \cite{LLNZ,gaiotto}.

In the simplest case of the sphere topology the Polyakov relation tells
us that $\partial S/\partial z_K = -\beta_K/2$ where $S$ is the
on-shell Liouville action, $z_K$ the position of the source and
$\beta_K$ the related accessory parameter.

The proof of Polyakov relation in presence of only parabolic
singularities was given in \cite{ZT} using fuchsian mapping techniques
and for the sphere in presence of both parabolic and elliptic
singularities using potential theory technique the proof was given in
\cite{CMS1,CMS2} and in \cite{TZ}.

In the present paper we shall extend such kind of relation to
higher genus surfaces showing that such a relation takes a different
meaning: not only it relates the change of the action under the motion
of the sources but also the change of the action under the change of
the moduli of the surface.
Here we give the proof of the relation in the case of the torus and in
the case of all hyperelliptic surface with an arbitrary number of
sources.

In the proof of Polyakov relation it is essential to exploit the
property of the accessory parameters to be real-analytic functions of
the position of the singularities and of the moduli of the surface.

This is not a trivial problem.  In paper \cite{kra} it was proven that
for the sphere the real-analytic dependence of the accessory parameters
on the position of the singularities holds everywhere in the
restricted case of parabolic and elliptic singularities of finite
order;  these are the singularities with strength $\eta = (1-1/n)/2$.

In (\cite{CMS1,CMS2}) it was proven for the sphere that the accessory
parameters are real-analytic functions of the positions of the
singularities and also on the strength $\eta$ of the singularities in
an everywhere dense open set for any collection of elliptic and$/$or
parabolic singularities without the restriction on the elliptic
singularities to be of finite order.

For the torus with one source a much stronger result was proven in
\cite{torusIII}, i.e. that the accessory parameter is a real-analytic
function of the coupling and of the modulus everywhere except for a
zero measure set. 

The proofs of the real-analyticity that we shall give in sections
\ref{counting},\ref{analyticity} rely heavily on the existence and the
uniqueness of the solution of the Liouville equation given the
strength, the positions of the singularities and the moduli of the
surface.

Starting from the papers of Picard \cite{picard}, which apply only to
elliptic singularities, there appeared various proof of the existence
and uniqueness of the solution of the 
Liouville equation \cite{poincare,lichtenstein,
mcowen,troyanov}.  The existence proofs are somewhat lengthy and
technical; on the other hand the uniqueness proof is rather
straightforward.

The proof of the almost-everywhere real-analytic property 
of the accessory parameter for the sphere with four sources and for
the torus with one source is
obtained by applying results and techniques related to analytic
varieties \cite{whitney,BM,GR} even though here we are in presence of
a problem of real-analytic varieties \cite{dangelo}.  This is dealt
with by the techniques of polarization, i.e. by doubling in the
intermediate steps of the proof the number of complex variables.

The paper is structured as follows. In section \ref{general} we give
the general discussion of the problem. In section \ref{actions} we
give the action on higher genus surfaces in two different coordinate
systems. In section \ref{auxiliary} we give the auxiliary differential
equation for the torus and all hyperelliptic surfaces with an
arbitrary number of sources.  In section \ref{counting} we give the
counting of the degrees of freedom of the parameters appearing in the
problem and write the implicit monodromy relations to which the
accessory parameters are subject. In section \ref{analyticity} we give
a shortened versions of the proof of the real-analyticity property of
the accessory parameters in an everywhere dense set in the general
case and in the case of the torus or of the four-point function on the
sphere, we give a shortened proof of the real-analyticity of the
accessory parameter everywhere except for a zero measure set.

In section \ref{proof} exploiting the results of the previous sections
we give the proof of the Polyakov relation for the sphere, the torus
and all hyperelliptic surfaces.  In section \ref{conclusions} we give
a short discussion of the results of the paper.

In the Appendix using results obtained in \cite{whitney} we derive the
analytic properties of zeros of Weierstrass polynomials which we need
in section \ref{analyticity}.

\section{General discussion}\label{general}

First we outline the semiclassical argument which leads to the
Polyakov relation. It will also serve to lay down the notation and fix
the normalization of the Liouville action which we shall choose as in
\cite{ZZ}.

The Liouville action, boundary terms apart, is given by
\begin{equation}
A_L=\frac{1}{\pi}\int(\partial_z\phi\partial_{\bar z}\phi+\pi\mu
  e^{2b\phi}) dz\wedge d\bar z\frac{i}{2}.
\end{equation}
with $z=x+iy$. The holomorphic energy momentum tensor is 
\begin{equation}
T_{zz}=T(z)=-(\partial_z\phi)^2+{\cal  Q}\partial_z^2\phi,~~~~~~~~
{\cal Q}=\frac{1}{b}+b 
\end{equation}
and for the vertex functions and their dimensions we have
\begin{equation}
V_\alpha(w)=e^{2\alpha\phi(w)},~~~~\Delta_\alpha=\alpha({\cal
  Q}-\alpha)
\end{equation}
\begin{equation}
\langle V_{\alpha_1}(w_1)\dots V_{\alpha_n}(w_n)\rangle=\int
V_{\alpha_1}(w_1)\dots V_{\alpha_n}(w_n)
e^{-A_L[\phi]} D[\phi]~.
\end{equation}
From the operator product expansion we have
\begin{equation}\label{ward}
T(z) V_\alpha(w)
=\frac{\Delta_\alpha}{(z-w)^2}V_\alpha(w)+\frac{1}{z-w}\partial_w
V_\alpha(w)+\dots
\end{equation}
To explore the semiclassical limit $b\rightarrow 0$ one sets
$\varphi=2b\phi$ and $\alpha=\frac{\eta}{b}$. The action and the
dimension $\Delta_\alpha$ become
\begin{equation}
A_L[\phi]=\frac{1}{b^2}S_L[\varphi]
~~~~~~~~\Delta_\alpha \approx \frac{1}{b^2}\eta(1-\eta)
\end{equation}
where, after performing a constant shift in $\varphi$
\begin{equation}
S_L[\varphi]=\frac{1}{2\pi}\int(\frac{1}{2} \partial_z\varphi
  \partial_{\bar z}\varphi+e^\varphi)dz\wedge d\bar z\frac{i}{2}
\end{equation}
and the energy momentum tensor becomes
\begin{equation}
T(z) \approx \frac{1}{b^2}\big[\frac{1}{2}\partial_z^2\varphi-
\frac{1}{4}(\partial_z \varphi)^2\big] =-\frac{1}{b^2}e^{\frac{\varphi}{2}}
\partial^2_z e^{-\frac{\varphi}{2}}~.
\end{equation}
Then in the semiclassical limit $b\rightarrow 0$
\begin{equation}
\langle V_{\alpha_1}(w_1)\dots V_{\alpha_n}(w_n)\rangle
= c~e^{-\frac{S^{cl}(w_1,\dots w_n)}{b^2}}
\end{equation}
where $S^{cl}(w_1,\dots w_n)$ is the classical action computed in
presence of the sources of strength $\eta_i$ at the points $w_i$.
As in the semiclassical limit the field is frozen on the classical
solution we also have
\begin{equation}
\langle T(z) V_{\alpha_1}(w_1)\dots \rangle=
\frac{c}{b^2}~(\frac{1}{2}\partial_z^2\varphi(z)-
\frac{1}{4}(\partial_z\varphi(z))^2)
e^{-\frac{S^{cl}(w_1,\dots w_n)}{b^2}}
\end{equation}
where
\begin{equation}
\frac{1}{2}\partial_z^2\varphi(z)-\frac{1}{4}(\partial_z\varphi(z))^2=
Q(z)=\sum_i\frac{1-\lambda_i^2}{4(z-w_i)^2}+\frac{\beta_i}{2(z-w_i)}~.
\end{equation}
Comparing with the result obtained using the 
operator product expansion (\ref{ward}) we have
\begin{equation}\label{polyakovrel}
\eta_i(1-\eta_i) =\frac{1-\lambda_i^2}{4},~~~~~~~~
\frac{\partial S^{cl}(w_1,\dots w_n}{\partial w_i})=-\frac{\beta_i}{2}~.
\end{equation}
As discussed in the introduction, proofs of (\ref{polyakovrel}) have
been given in \cite{CMS1,CMS2,ZT,TZ} for the topology of the sphere.

In the case of the torus we have two simple representations of the
manifold. One is the quotient of the complex $z$-plane by the group of
discrete translations with generators $2\omega_1$, $2\omega_2$ and the
other is the Weierstrass representation via the variable $u=\wp(z)$.
One can use as parameter classifying the torus the modulus
$\tau=\omega_2/\omega_1$ as done in \cite{torusIII}, but both for the
torus and for higher genus it will be simpler to use the position of
the branch points of the two sheet representation of the elliptic or
hyperelliptic surface.

For $g=2$ the analogue of the $\wp$ function was given in
\cite{komori}. For an approach to the $g=3$ problem see
\cite{eilbeck}.

On the other hand we know that for any genus $g\geq 2$ we can
represent the Riemann surface as the quotient of the $z$ upper
half-plane by a fuchsian group i.e. by a standard fundamental
curvilinear polygon \cite{FK}. Elliptic and hyperelliptic surfaces of
any genus can be represented by a two sheet cut $u$-plane.  Even
though the transformation between the two representation is not known
explicitly except for $g=1$ and $g=2$, we find in section
\ref{actions} general properties of the Jacobian relating the
$z$-representation with the two sheet $u$-representation of the
hyperelliptic surface. This will be sufficient to relate the actions
in the two representations.

Accessory parameters appear through the auxiliary ordinary differential
equation associated with the Liouville problem. For elliptic and hyperelliptic
surfaces they are in number $n+2g+1$ for $g\leq 2$, $n$ being the
number of sources and $g$ the genus of the surface and for $g\geq 3$
they are in number $3g+n-1$. However we have relations among them, the
fuchsian relations, with the final result that for any $g$ we have
$n+3g-3$ independent accessory parameters. This is true also in the
general case of non-hyperelliptic surfaces \cite{kra}.

As we mentioned an hyperelliptic surface can be
represented in several form. It will turn out that the simplest choice
is to use for the moduli the locations $u_l$ of the branch points
of the two sheet representation 
of the manifold; for this choice the Polyakov relation takes the form
\begin{equation}
\frac{\partial S}{\partial u_l} = -B_l -\frac{1}{8}(\partial_s \varphi_M)^2
\end{equation}
where $B_l$ is the accessory parameter at the branch point $u_l$ and
$\varphi_M$ is the regular part of the Liouville field at the singularity.

In the process of taking the derivative of the classical action with
respect to the locations of the sources $u_K$ or to the moduli
$u_l$ , one has to keep in mind that the classical solutions depend on
such positions and, through the auxiliary equation, also on the values
of the $\beta$'s and of a real weight parameter $\kappa$
which are fixed by the monodromy conditions. Here is
where the real-analyticity of the $\beta$'s as functions of the
$u_K,u_l$ enters the problem.

The normalization of the action $S$ we use in this paper is the one
adopted in \cite{ZZ}; it is related to the one used in
\cite{torusIII,torusI,torusII} which we call $S_T$ by $S_T=2\pi S$ and
to the one used in \cite{CMS1,CMS2} and in \cite{ZT,TZ} which we call
$S_{CMS}$ by $S_{CMS} = 4\pi S$.

\section{The action on higher genus surfaces}\label{actions}

For completeness we start recalling the action on a surface with the
topology of the sphere.

\bigskip

$g=0$. 

The sphere is described by $C\cup \infty$ and the action is given
by
\begin{eqnarray}\label{actionSphere}
S &=& \frac{1}{2\pi}\int_{D_\epsilon} (\frac{1}{2}
\partial\phi\wedge\bar \partial\phi+e^\phi
dz\wedge \bar dz)\frac{i}{2}\nonumber \\
&-&\frac{\eta_K}{4\pi i}\oint_{\epsilon_K}\phi(\frac{dz}{z-z_K}-
\frac{d\bar z}{\bar z-\bar z_K})-\eta^2_K\log\epsilon^2_K \nonumber\\
&+&\frac{1}{4\pi i}\oint_{R}\phi(\frac{dz}{z}
-\frac{d\bar z}{\bar z})+\log R^2
\end{eqnarray}
in the limit $\epsilon_K\rightarrow 0$, $R\rightarrow\infty$
where $D_\epsilon$ is the disk of radius $R$ in the complex plane from which
disks of radius $\epsilon_K$ around $z_K$ have been removed. We use
the notation
$\partial f \equiv \partial_z f dz,~\bar\partial f \equiv 
\partial_{\bar z} f d\bar z$.
Variation of such an action, with $\phi$ satisfying at $z_K$
the boundary conditions 
\begin{equation}\label{bcelliptic}
\phi(z)= -2\eta_K\log(z-z_K)(\bar z-\bar z_K)+ r_K
\end{equation}
and at $z=\infty$ the boundary condition 
\begin{equation}
\phi(z)= -2\log z\bar z + r_\infty
\end{equation}
where $r_K,~r_\infty$ are bounded continuous functions,
gives rise to the Liouville equation 
\begin{equation}\label{simpleLiouville}
-\partial_z\partial_{\bar z}\phi+e^\phi =0
\end{equation}
in $C\backslash \{u_K\}$.
We shall write for the solution of Liouville equation
\begin{equation}
r_K = X_K+ o(z-z_K),~~~~~~~~r_\infty = X_\infty +o(\frac{1}{z})~.
\end{equation}
The $\eta_K$ are subject the restrictions $\eta_K<\frac{1}{2}$
(local finiteness of the area) and to the topological restriction
$\sum_K 2\eta_K>2(1-g) =\chi =2$, where $g$ is the genus and $\chi$
the Euler characteristic.

In the case of parabolic singularities
the behavior of the field at the singularities is
\begin{equation}\label{bcparabolic}
\phi(z)= -\log(z-z_P)(\bar z-\bar z_P)
-\log(\log(z-z_P)(\bar z-\bar z_P))^2 +r_P
\end{equation}
and in the action (\ref{actionSphere}) and in the previous topological
relation $\eta_K$ has to be replaced by $\frac{1}{2}$.

\bigskip

$g=1$ 

The torus is described by the quotient of the
complex plane by the discrete translation group with generators
$2\omega_1,2\omega_2$
and Liouville equation is given by eq.(\ref{simpleLiouville})
with periodic boundary conditions in $z$ and 
$\phi$ behaving as eq.(\ref{bcelliptic},\ref{bcparabolic}) 
at the singularities and $\sum_K 2\eta_K +\sum_P 1 >2(1-g)=\chi =0$.

In such a $z$-representation the action is given by
\begin{equation}\label{SzactionTorus}
S_z=\frac{1}{2\pi}\int_{T}(\frac{1}{2}
\partial\phi\wedge\bar \partial\phi+e^\phi dz\wedge d\bar z)\frac{i}{2}-
\frac{\eta_K}{4\pi i}\oint_{\epsilon_K}\phi(\frac{dz}{z-z_K}-
\frac{dz}{z-z_K})-\eta_K^2\log\epsilon^2_K
\end{equation}
where the index $K$ runs on the sources. Due to the periodic boundary
conditions on $\phi$ we have no boundary terms.
Working with periodic boundary conditions is not very simple.
It is useful to go over to the Weierstrass representation of
the torus given by the equation
\begin{equation}
w^2=4(v-e_1)(v-e_2)(v-e_3),~~~~~~~~e_1+e_2+e_3=0~.
\end{equation}
Actually to connect to the general hyperelliptic case it is useful 
to maintain a more general formalism in which 
$v= u-(u_1+u_2+u_3)/3$ and $e_l= u_l-(u_1+u_2+u_3)/3$ so that the
equation for the manifold becomes
\begin{equation}\label{W-uequation}
w^2=4(u-u_1)(u-u_2)(u-u_3)
\end{equation}
and
\begin{equation}
\wp(z) = v= u-\frac{u_1+u_2+u_3}{3}~.
\end{equation}
From the well known differential equation satisfied by $\wp(z)$
\begin{equation}
(\wp'(z))^2 = 4(\wp(z)-e_1)(\wp(z)-e_2)(\wp(z)-e_3)
\end{equation}
we have 
\begin{equation}
J=\frac{dz}{du}=\frac{1}{\sqrt{4(u-u_1)(u-u_2)(u-u_3)}}~,
\end{equation}
\begin{equation}
z=\int_\infty^u\frac{du}{\sqrt{4(u-u_1)(u-u_2)(u-u_3)}}~.
\end{equation}
A point $p$ of the surface is given by the couple of numbers $(u,w)$
where $w$ satisfies eq.(\ref{W-uequation}) and thus it can assume two values.

For the torus we have for the half-periods
\begin{equation}
\omega_1 = \frac{1}{\sqrt{u_1-u_2}} K\bigg(\sqrt{\frac{u_3-u_2}{u_1-u_2}}\bigg)
\end{equation}
\begin{equation}
\omega_2 =\frac{i}{\sqrt{u_1-u_2}}K\bigg(\sqrt{\frac{u_1-u_3}{u_1-u_2}}\bigg)
\end{equation}
and the modulus is $\tau=\omega_2/\omega_1$. In studying the dependence
of the action on the moduli, one can use for the torus $\tau$ as done
in \cite{torusIII,torusI,torusII}. 
On the other hand both for the torus and for the
general hyperelliptic surface it is simpler to classify the surfaces in
terms of the positions of the branch points of the map from the
fundamental standard polygon to the two sheeted $u$-plane.

Due to the invariance of
the area i.e. $e^\varphi du\wedge d\bar u=e^\phi dz\wedge d\bar z$,
in the $u$-representation the field is given by
\begin{equation}\label{phivarphitorus}
\varphi(u) =\phi(z)+\log J\bar J,~~~~~~~~J=\frac{dz}{du}~.
\end{equation}
From the behavior of the field $\phi(z)$ at the sources
\begin{equation}
\phi(z) = -2\eta_K\log(z-z_K)(\bar z-\bar z_K)+ X_K+ o(z-z_K)
\end{equation}
we have that the behavior of $\varphi(u)$ at the sources is
\begin{equation}\label{varphiKbc}
\varphi(u) = -2\eta_K\log(u-u_K)(\bar u-\bar u_K)+ X^u_K+ o(u-u_K)
\end{equation}
with
\begin{equation}\label{XXurelation}
X^u_K = X_K-(1-2\eta_K)\log|4(u_K-u_1)(u_K-u_2)(u_K-u_3)|~.
\end{equation}
At $u=\infty$ being $\phi(0)$ finite we have
\begin{equation}\label{behaviortorusinfinity}
\varphi(u) = \phi(0) -\log 4-\frac{3}{2}\log u\bar u+o(\frac{1}{u})~.
\end{equation}
In the following we shall use the convention to denote the dynamical
singularities i.e. the sources by $u_K$ with upper case index, while the
kinematical singularities describing the Riemann surface in the
$u$-representation will be denoted by $u_l$, with lower case index.

The action in the $u$-representation $S_u$ taking into account
the behavior (\ref{behaviortorusinfinity}) is given by
\begin{eqnarray}\label{SuactionTorus}
S_u&=& 
\frac{1}{2\pi}\int_{D_\varepsilon}(\frac{1}{2}\partial\varphi\wedge\bar
\partial\varphi+ e^\varphi du\wedge d\bar u)
\frac{i}{2}\nonumber\\
&-&\frac{\eta_K}{4\pi i}\oint_{\varepsilon_K} \varphi(\frac{du}{u-u_K} -\frac{d\bar
  u}{\bar u-\bar u_K})-\eta_K^2\log\varepsilon_K^2\nonumber\\
&-&\frac{1}{16\pi i}\oint^d_{\varepsilon_l}\varphi(\frac{du}{u-u_l} 
-\frac{d\bar  u}{\bar u-\bar u_l}) 
-\frac{1}{8}\log\varepsilon_l^2\nonumber\\
&+&\frac{1}{8\pi i}\frac{3}{2}\oint^d_{R_u}\varphi(\frac{du}{u} 
-\frac{d\bar  u}{\bar u}) 
+\frac{1}{2}\big(\frac{3}{2}\big)^2\log R_u^2
\end{eqnarray}
where $D_\varepsilon$ is the double sheeted plane 
and the index $d$ on the contour integrals means that a double turn 
has to be taken around the kinematical
singularities $u_l$, $l=1,2,3$ and at $\infty$ in order to come back
to the starting point.

For the actions $S_z$ and $S_u$ the general relations 
\cite{ZZ,CMS1,CMS2} hold
\begin{equation}\label{Xrelation}
\frac{\partial S_z}{\partial \eta_K} =-X_K~, ~~~~~~~~
\frac{\partial S_u}{\partial \eta_K} =-X^u_K 
\end{equation}
which are easily proven from the form 
(\ref{actionSphere},\ref{SzactionTorus},\ref{SuactionTorus})
of the actions.

The relation between the two actions is obtained by replacing in
${ S}_z$, $\phi$ in terms of $\varphi$ as given by equation 
(\ref{phivarphitorus}). We find 
\begin{equation}\label{actiontorusz-actiontorusu}
S_z=S_u-\sum_K\eta_K(1-\eta_K) \log[4
|u_K-u_1||u_K-u_2||u_K-u_3|]-\frac{1}{2}\log[
|u_1-u_2||u_2-u_3||u_3-u_1|]~.
\end{equation}
We notice that eq.(\ref{actiontorusz-actiontorusu}) is consistent 
with the general relation (\ref{Xrelation}) combined with 
(\ref{XXurelation}). 

The difference between the two actions is of dynamical nature as
it involves the source strengths $\eta_K$.

In the above equation one recognizes the classical dimensions of the
sources $\eta_K(1-\eta_K)$ multiplied by the logarithm of the Jacobian
of the transformation. 

\bigskip

$g\geq 2$

We know that a compact Riemann surface of genus $g\geq 2$ can be
represented by a standard fundamental domain of the complex upper
half-plane. Such a domain is a curvilinear $4g$-gon which is the
analog of the parallelogram $T$ belonging to $C$ which describes the
torus. 
Surfaces of genus $g=2$ are all hyperelliptic. For these, $g=2$ Komori
\cite{komori} gave an explicit representation in terms of the analogue
of the Weierstrass function $\wp(z)$, which we shall call $h(z)$, as
the ratio of two 6-forms
\begin{equation}
h(z)=\frac{f(z)}{g(z)}
\end{equation}
where $f(z)$ and $g(z)$ are explicitly written in terms of Poincar\'e
series on a fuchsian group $G$.
Then we have the representation 
\begin{equation}
w^2 = 4(u-h(z_1))(u-h(z_2))(u-h(z_3))(u-h(z_4))(u-h(z_5))
\end{equation}
with $u=h(z)$.
\begin{equation}
f(z) = \sum_{\gamma\in G}\frac{1}{\gamma z-p_6}P(\gamma z)\gamma'(z)^3
\end{equation}
\begin{equation}
g(z) = \sum_{\gamma\in G}P(\gamma z)\gamma'(z)^3
\end{equation}
and $f(z)$ has simple poles on the orbit of the point $p_6$.
$P(z)$ is a properly constructed rational function of $z$ holomorphic
in the upper half-plane \cite{komori}.

In the following we shall enucleate the general features of the
transformation between the $z$ and $u$ coordinates for hyperelliptic
surfaces of any genus. This we be sufficient to relate $S_z$ with
$S_u$.

The structure of the Jacobian of the transformation 
\begin{equation}
J=\frac{dz}{du}
\end{equation}
can be extracted as follows.
The surface is described by
\begin{equation}
w^2 = 4(u-u_1)\dots(u-u_{2g+1})~.
\end{equation}

$(u,w)$ is a faithful representation of our Riemann surface and thus
to each such point there correspond a point in the standard
fundamental polygon in the $z$-upper-half-plane; $z$ is a locally
conformal (analytic invertible) representation of the Riemann surface.

In a domain around a point of $M$, described by 
$(u,w)$ with $u\neq u_l$, $M$ is represented by $(u,w)$ 
with $w$ a determination of $\sqrt{4(u-u_1)\dots(u-u_{2g+1})}$.
In a domain around the point of $M$, described by 
$(u_l,0)$, $M$ is faithfully represented by $w$. 
In the first case $z$ is an
analytic (locally invertible) function of $u$, while in the second case
we have 
\begin{equation}
z-z_l = w f_l(w)
\end{equation}
with $f_l$ analytic and $f_l(0)\neq 0$ and $u$ function of $w$
according to 
\begin{equation}
u-u_l = \frac{w^2}{4(u-u_1)\dots\{(u-u_l)\}\dots(u-u_{2g+1})}
\end{equation}
where the term in $\{\}$ has to be removed.

The Jacobian is given by 
\begin{eqnarray}
J= \frac{dz}{du}=\frac{dz}{dw}
\frac{dw}{du} =(f_l(w)+wf_l'(w))
\frac{(u-u_1)\dots\{(u-u_l)\}\dots(u-u_{2g+1})+O(u-u_l)}{\sqrt{(u-u_1)\dots
(u-u_{2g+1})}} \nonumber \\
=2 (f_l(w)+wf_l'(w))
\frac{(u-u_1)\dots\{(u-u_l)\}\dots(u-u_{2g+1})+O(u-u_l)}{w} 
\end{eqnarray}
so that 
\begin{eqnarray}
\log J &=& -\frac{1}{2}\log(u-u_l)+\log f_l(0)+
\frac{1}{2}\log[(u_l-u_1)\dots\{(u_l-u_l)\}\dots(u_l-u_{2g+1})]\nonumber\\
&+&\nonumber 
O(\sqrt{u-u_l})\nonumber \\
&\equiv&
 -\frac{1}{2}\log(u-u_l)+j_l+ O(\sqrt{u-u_l})
\end{eqnarray}
where
\begin{equation}
j_l = \log f_l(0)+\frac{1}{2}\log[(u_l-u_1)\dots\{(u_l-u_l)\}
\dots(u_l-u_{2g+1})]~.
\end{equation}

With regard to the fields we have:

At $u_K$ from
\begin{equation}
\phi(z) = -2\eta_K\log(z-z_K)(\bar z-\bar z_K)+X_K +o(z-z_K)
\end{equation}
we deduce 
\begin{eqnarray}\label{varphiKbc}
&&\varphi(u) = -2\eta_K \log(u-u_K)(\bar u-\bar u_K)+(1-2\eta_K)\log
J_K\bar J_K+X_K +O(u-u_K)=\nonumber\\
&&-2\eta_K \log(u-u_K)(\bar u-\bar u_K)+X^u_K +O(u-u_K)
\end{eqnarray}
with
\begin{equation}
X^u_K =X_K+(1-2\eta_K)\log J_K\bar J_K~.
\end{equation}

At $u_l$ we have 
\begin{eqnarray}\label{varphilbc}
&&\varphi(u) = -\frac{1}{2} \log(u-u_l)(\bar u-\bar u_l) +j_l
+\bar j_l+\phi(z_l)+O(\sqrt{u-u_l})=\nonumber\\
&&-\frac{1}{2} \log(u-u_l)(\bar u-\bar u_l) +X_l^u+O(\sqrt{u-u_l})
\end{eqnarray}
with 
\begin{equation}
X^u_l =j_l+\bar j_l+\phi(z_l)~.
\end{equation}
We shall also need the behavior of $\varphi$ at infinity in $u$. The
local uniformizing variable in the $u$ cut-plane at infinity is
$v$ with $v^2=1/u$. Then being $z$ in a neighborhood of $z_\infty$ 
(i.e. of the point which is projected to $u=\infty$) a regular
representation of the manifold we have
\begin{equation}
z-z_\infty = \alpha_\infty v +O(v^2)~.
\end{equation}
Thus 
\begin{equation}
J
=\frac{dz}{du}=\frac{dz}{dv}\frac{dv}{du}=-\frac{\alpha_\infty}{2} 
u^{-\frac{3}{2}}(1+O(v))
\end{equation}
and
\begin{equation}
\log J =-\frac{3}{2} \log u +j_\infty,~~~~~~~~j_\infty 
= \log(-\frac{\alpha_\infty}{2})~.
\end{equation}
Then as $\varphi =\phi +\log J\bar J$ we have
\begin{equation}\label{varphiinfty}
\varphi(u) = \phi(z_\infty)-\frac{3}{2}\log u \bar u +j_\infty+\bar j_\infty
=-\frac{3}{2}\log u \bar u+X^u_\infty,~~~~~~~~X^u_\infty=
\phi(z_\infty)+j_\infty+\bar j_\infty~.
\end{equation}

Integrating
\begin{equation}
\partial_u\partial_{\bar u}\varphi = e^\varphi,~~~~~~~~
e^\varphi du\wedge d\bar u \frac{i}{2}>0
\end{equation}
we obtain the topological inequality for the source strengths $\eta_K$, the
number of parabolic singularities and the genus $g$
\begin{eqnarray}
&&0<\frac{i}{2}\bigg(-\oint_{u_K}\bar\partial\varphi-
\oint_{u_l}^d\bar\partial\varphi+\oint_\infty^d\bar\partial\varphi\bigg)=
\pi\big(\sum_K 2\eta_K-3+\sum_P 1+\sum_{l=1}^{2g+1} 1\big)\nonumber\\
&=&\pi\big(\sum_K 2\eta_K+\sum_P 1+2(g-1)\big)~.
\end{eqnarray}

For $g\geq 2$ the compact Riemann
surface is represented by the quotient of the upper $z$-plane by a Fuchsian
group \cite{FK}.  We refer to a standard fundamental polygon $D_z$. It is a
curvilinear polygon with $4g$ sides lying in the upper $z$ plane with
all vertices identified. The sides lie in the order $A_1 B_1 A^{-1}_1
B^{-1}_1 \dots A^{-1}_g B^{-1}_g$. There exist one and only one
element $\Gamma^A_j$ of the fuchsian group which maps $A_j$ into
$A^{-1}_j$ and one and only one element $\Gamma^B_j$ of the
fuchsian group which maps $B_j$ into $B^{-1}_j$ \cite{FK}.

The side $A_j$ is identified with the side $A^{-1}_j$ and when one
runs along the perimeter of the $4g$-gon the image of $A_j$, $\Gamma_j
A_j = A^{-1}_j$ is is traveled in the opposite direction as
$A_j$. Thus the contour $A_j A_j^{-1}$ is a closed loop on the Riemann
surface.

The action in the $z$-representation is given by 
\begin{eqnarray}\label{SzactionGen}
&&S_z= \frac{1}{2\pi}\int_{D_z}
\big(\frac{1}{2}\partial\phi\wedge\bar \partial\phi+e^\phi dz\wedge d\bar z\big
)\frac{i}{2}
-\frac{\eta_K}{4\pi i}\oint_{\epsilon_K} \phi\big(\frac{dz}{z-z_K}-\frac{d\bar
  z}{\bar z-\bar z_K}\big)-\eta_K^2 \log\epsilon^2_K\nonumber\\
&&+\frac{i}{8\pi}\int_{A_j} \phi(\bar\partial\log \bar s^A_{j}-\partial\log
s^A_{j})+
\frac{i}{8\pi}\int_{B_j} \phi(\bar\partial\log \bar s^B_{j}-\partial\log s^B_{j})~.
\end{eqnarray} 
The one dimensional integrals in the last line of the above equation
are  boundary terms and they are present 
due to the fact that $\phi$ is not a scalar but a conformal
field, i.e. the periodic boundary conditions are on $e^\phi dz\wedge
d\bar z$ and not on $\phi$. 
The $s$ are given by $s=dz/dz'$. If the transformation
$\Gamma$ which relates two identified sides $A$ and $A^{-1}$ is given
by
\begin{equation}
\Gamma(z)=z'=\frac{az+b}{cz+d}
\end{equation}
we have $s=(cz+d)^2$ and $\partial \log s=2c~(dz)/(cz+d)$.

In addition $D_z$ excludes small circles of radius $\epsilon_K$ around the
sources $z_K$ and, as an intermediate step, small circles around $z_l$, being
$z_l$ the images of the $u_l$ 
and around $z_\infty$, the image of $u=\infty$.
The field dependent boundary terms of the last line in
(\ref{SzactionGen}) are absent for
the torus due to the linear nature of the $\Gamma$ is such a case.

Substituting in the above equation $\phi= \varphi-\log J\bar J$ and
using the information on $J$ derived previously in this section, we
obtain the relation between the action $S_z$ and the action in the
$u$-representation, $S_u$ 
\begin{eqnarray}\label{SzactionGenSuactionGen}
&&S_z=S_u+\eta_K(1-\eta_K)
\log(J_K\bar J_K)+
\frac{1}{4}(j_l+\bar j_l)-\frac{3}{4}(j_\infty+\bar j_\infty)\nonumber\\
&&-\frac{i}{8\pi}\int_{A_j}\log(J\bar J) (\bar\partial\log \bar s^A_j-
\partial\log s^A_j)
-\frac{i}{8\pi}\int_{B_j}\log(J\bar J) (\bar\partial\log\bar s^B_j-
\partial\log s^B_j)\nonumber\\
&-&\frac{i}{8\pi}\oint\log(J\bar J)\frac{d\log J}{dz}dz
\end{eqnarray}
where the last term is the contour integral along the boundary of the
standard fundamental domain. Sums over $K$, $l$ and $j$ are understood.

$S_u$ is given by
\begin{eqnarray}\label{SuactionGen}
S_u&=&\frac{1}{2\pi}\int_{D_u}(\frac{1}{2}\partial\varphi
\wedge\bar
\partial\varphi+e^\varphi du\wedge d\bar u)\frac{i}{2}
-\frac{\eta_K}{4\pi i}\oint_{\varepsilon_K}\varphi(\frac{du}{u-u_K}-
\frac{d\bar u}{\bar u-\bar u_K})-\eta_K^2\log\varepsilon^2_K\nonumber\\
&-&\frac{1}{16\pi i}\oint^d_{\varepsilon_l}\varphi(\frac{du}{u-u_l}-
\frac{d\bar u}{\bar u-\bar
  u_l})-\frac{1}{8}\log\varepsilon^2_l\nonumber\\
&+&\frac{1}{8\pi i}\frac{3}{2}\oint^d_{R_u}\varphi(\frac{du}{u}-
\frac{d\bar u}{\bar u})+\frac{1}{2}\big(\frac{3}{2}\big)^2\log R^2_u~.
\end{eqnarray}

Such action with the boundary conditions
(\ref{varphiKbc},\ref{varphilbc},\ref{varphiinfty}) is finite.
Its variation, again with the boundary conditions
(\ref{varphiKbc},\ref{varphilbc},\ref{varphiinfty}), 
gives rise to the equation of motion
\begin{equation}\label{pureliouvilleu}
-\partial\bar\partial\varphi + e^\varphi du\wedge d\bar u =0
\end{equation}
in the two-sheeted cut $u$-plane with the singular points
$(u_K,w_K),(u_l,0)$ removed.
The field dependent boundary terms
appearing in the last line of eq.(\ref{SzactionGen}) are canceled when
performing the above described transition from $\phi$ to $\varphi$.

The general relations (\ref{Xrelation}) hold also for the actions
(\ref{SzactionGen},\ref{SuactionGen}) and they are consistent with the
term $\eta_K(1-\eta_K)\log(J_K\bar J_K)$ appearing in 
eq.(\ref{SzactionGenSuactionGen}) and the relation
\begin{equation}
X^u_K-X_K = (1-2\eta_K)\log J_K\bar J_K~.
\end{equation}

\section{The auxiliary differential equation}\label{auxiliary}

Given the field $\phi(z)$ we know that in virtue of the Liouville equation
\begin{equation}
e^{\frac{\phi}{2}} \partial_z^2 e^{-\frac{\phi}{2}}\equiv -Q_z(z)
\end{equation}
is analytic in $z$ except for first and second order poles. 
Under a change of coordinates e.g. from $z$ to $u$,
the $Q$ transforms as follows
\begin{equation}\label{schwarztranformation}
Q_u(u)=Q_z(z)\bigg(\frac{dz}{du}\bigg)^2-\{z,u\}
\end{equation}
where $\{z,u\}$ is the Schwarz derivative
\begin{equation}
\{z,u\}=(\frac{dz}{du})^\frac{1}{2}\frac{d^2}{du^2}
(\frac{dz}{du})^{-\frac{1}{2}}~.
\end{equation}

Given the differential equation
\begin{equation}\label{auxiliaryeq}
f''(u)+Q_u(u)f(u)=0
\end{equation}
we know (see e.g. \cite{CMS2,torusI,torusII}) 
that the conformal factor can be expressed as
\begin{equation}\label{expvarphi}
e^{\varphi(u)} = \frac{2 w_{12} \bar w_{12}}
{[\kappa^{-2}f_1(u)\bar f_1(\bar u)-\kappa^2 f_2(u)\bar f_2(\bar u)]^2}
\end{equation}
where $f_1,f_2$ are properly chosen solutions of
eq.(\ref{auxiliaryeq})
and $w_{12}$ their Wronskian.

The accessory parameters appear in the ordinary differential
equation (\ref{auxiliaryeq}) associated with the
Liouville problem. We give in the following the structure of the
differential equation in canonical form.

For the torus with a single source at $z=z_1$ we have the equation \cite{keen}
\begin{equation}
f''(z)+\epsilon(\wp(z-z_1)+\beta)f(z)=0
\end{equation}
but we are interested in the case with $n$ sources and of the general 
hyperelliptic surface for which the $u$ representation is simpler.
The general form of $Q_u(u)$ for any hyperelliptic surface with $n$
sources is
\begin{eqnarray}\label{generalQu}
Q_u&=&\frac{3}{16}\bigg(\frac{1}{(u-u_1)^2}+\dots+\frac{1}{(u-u_{2g+1})^2}
\bigg)\nonumber\\
&+&
\frac{\beta_1}{2(u-u_1)}+\dots+\frac{\beta_{2g+1}}{2(u-u_{2g+1})}\nonumber\\
&+&
\sum_K\bigg(\epsilon_K\frac{(w+w_K)^2}{4(u-u_K)^2w^2}+
\frac{\beta_K(w+w_K)}{4(u-u_K)w}\bigg)\nonumber\\
&+& \frac{\beta^{(0)}_w}{w}+\dots +\frac{u^{g-3}\beta^{(g-3)}_w}{w}
\end{eqnarray}

with $\epsilon_K=(1-\lambda^2_K)/4 = \eta_K(1-\eta_K)$,
$w=\sqrt{4(u-u_1)\dots(u-u_{2g+1})}$ and\break
$w_K=\sqrt{4(u_K-u_1)\dots(u_K-u_{2g+1})}$ 
where the last line is present only for $g\geq 3$.

The structure of $Q_u$ has the following origin. To each kinematical
singularity at $u_l$ with $l=1\dots 2g+1$ there corresponds an
accessory parameter $\beta_l$. To each dynamical singularity at
$(u_K,w_K)$ there corresponds an accessory parameter $\beta_K$.  The
factors $(w+w_K)/(2w)$ and their squares project the singularity on
the correct sheet. With regard to the kinematical singularities $u_l$
we notice that to the total accessory parameter contribute not only
$\beta_l$ but also the terms $1/w^2$, as explicitly given in section
\ref{proof}.

The function $\frac{1}{w}$ 
does not introduce a singularity at $w=0$ as seen going over to the
local uniformizing variable $s^2 =u-u_l$ and computing the related $Q_s$.  

At infinity the uniformizing variable is $v$ given by $u=v^{-2}$
and we have for the related $Q_v$
\begin{equation}
Q_v = Q_u \bigg(\frac{du}{dv}\bigg)^2-\{u,v\},~~~~~~~~\frac{du}{dv}=-2
v^{-3},~~~~\{u,v\}=\frac{3}{4v^2}~.
\end{equation}
The terms of the last line in eq.(\ref{generalQu}) are allowed provided, when
combined with the other contributions, leave the $Q_v$ free of
singularity at $v=0$.
The contribution of the term $\frac{u^p}{w}$ to $Q_v$ is
\begin{equation}
\frac{u^p}{w}\bigg(\frac{du}{dv}\bigg)^2 \sim v^{2g-2p-5}~~~~
{\rm for}~~v\approx 0
\end{equation} 
and thus they are consistent with the regularity at infinity 
only for $2g-2p-5 \geq 0$ and this
happens for $g\geq 3$. 
The $\beta$'s appearing in eq.(\ref{generalQu}) 
are subject to the conditions of absence
of sources at infinity. The structure is special for $g=1$ and $g=2$
while it becomes systematic for $g\geq 3$.

Explicitly:

For $g=1$ the number $\beta$'s is $n+3$  and we have the
three conditions from the regularity at infinity.
\begin{equation}
2(\beta_1+\beta_2+\beta_3)+\sum_K \beta_K=0
\end{equation}
\begin{equation}
\frac{3}{2}+ \sum_K(\epsilon_K 
+u_K\beta_K)+ 2(u_1\beta_1+u_2\beta_2+u_3\beta_3)=0
\end{equation}
\begin{equation}
\sum_K w_K \beta_K=0
\end{equation}
thus leaving $n$ free $\beta$'s.

For $g=2$ the number of $\beta$'s is $n+5$ and we have 
only two conditions given by
\begin{equation}
2(\beta_1+\beta_2+\beta_3+\beta_4+\beta_5)+\sum_K \beta_K=0
\end{equation}
\begin{equation}
3+ \sum_K(\epsilon_K +u_K\beta_K)+
2(u_1\beta_1+u_2\beta_2+u_3\beta_4+u_4\beta_3+u_5\beta_5)=0
\end{equation}
leaving us with $n+3$ $\beta$'s.

For $g=3$ the number of $\beta$'s is $8+n$ and we 
have the two  conditions
\begin{equation}
2(\beta_1+\dots+\beta_7)+\sum_K \beta_K=0
\end{equation}
\begin{equation}
\frac{9}{2}+ \sum_K(\epsilon_K +u_K\beta_K)+
2(u_1\beta_1+\dots+u_7\beta_7)=0
\end{equation}
which leaves us with $n+6$ independent $\beta$'s.

From now on, increasing the genus by one we introduce three more
$\beta$'s while the constraints remain always two. Thus we have
recovered from the study of $Q_u$ for the number of independent
accessory parameters the general formula $3g-3+n$.

From expression
(\ref{expvarphi}) we have in a neighborhood of an elliptic
singularity $u_K$
with $\zeta = u-u_K$
\begin{equation}\label{behavioruK}  
\varphi = -2\eta_K \log(\zeta\bar \zeta)-2\log \big[f(\zeta)\bar f(\bar
  \zeta)-\kappa^4(\zeta\bar\zeta)^{\lambda_K} g(\zeta)\bar g(\bar
  \zeta)\big]+2\log|\kappa|^2+\log(2 w_{12}\bar w_{12})
\end{equation}
where $f(\zeta)$ and $g(\zeta)$ are given
by a locally convergent power expansions.
Around parabolic singularities we have the expression \cite{CMS1,CMS2} 
\begin{equation}\label{behavioruP}
\varphi = -\log\zeta\bar\zeta - \log\log^2(\zeta\bar\zeta)-
2\log\bigg[g(\zeta)\bar g(\bar\zeta)+\frac{f(\zeta)\bar g(\bar\zeta)
+\bar f(\bar \zeta) g(\zeta)}{\log\frac{\zeta\bar\zeta}{\kappa^4}}\bigg]
+{\rm const}~.
\end{equation}  
Around a kinematical singularity $u_l$ the local uniformizing variable
is $s$ with $s^2= u-u_l$; in eq.(\ref{behavioruK}) $\eta_K$ 
has to be replaced by $1/4$ and $f$ and $g$ become power expansions in
$s$. The detailed form is given in section \ref{proof}.

At infinity we have
for the sphere
$\phi= -2\log z\bar z + h(\frac{1}{z}, \frac{1}{\bar z})$ and for
higher genus $\varphi= -\frac{3}{2}\log u\bar u + h(\frac{1}{u},
\frac{1}{\bar u})$ with $h$ analytic function in the two variables.
This information can be used to give a very simple proof of the
uniqueness of the solution of Liouville equation on the sphere, the
torus and
hyperelliptic surfaces of any genus in presence of any collection of
elliptic and parabolic singularities.

Consider two solutions $\varphi_1$ and $\varphi_2$ of
eq.(\ref{pureliouvilleu}) satisfying the above boundary conditions.
Then we have
\begin{eqnarray}
0&\leq& \frac{i}{2}\int \partial(\varphi_2-\varphi_1)  
\bar\partial(\varphi_2-\varphi_1)= 
\frac{i}{2}\oint (\varphi_2-\varphi_1) \bar\partial(\varphi_2-\varphi_1)-
\frac{i}{2}\int (\varphi_2-\varphi_1)\partial\bar
\partial(\varphi_2-\varphi_1)\nonumber\\
&=&
0-\int (\varphi_2-\varphi_1)
(e^{\varphi_2}-e^{\varphi_1})d^2u~.
\end{eqnarray}  
The contour integral is around the singularities $u_K$, $u_l$  
and at infinity and due to the behavior of $\varphi_2-\varphi_1$ 
it vanishes.
Thus we have
$\varphi_2=\varphi_1$.  Picard's uniqueness argument \cite{picard} 
is more complicated
because he did not use the information about the non leading terms
appearing in eqs.(\ref{behavioruK},\ref{behavioruP}) 
provided by the
auxiliary differential equation (\ref{auxiliaryeq}).

\section{Realization of the $SU(1,1)$ monodromies}\label{counting}

The existence and uniqueness proofs for the solutions of Liouville
equations \cite{picard,poincare,lichtenstein,mcowen,troyanov}
give us information on the accessory parameters. In fact
given the solution $\varphi(u)$ we have
\begin{equation}\label{Qfromphi}
e^{\phi/2} \partial^2_u e^{-\phi/2} = - Q(u)
\end{equation}
The accessory parameters $\beta$ appear explititely in the expression
of $Q(u)$ (\ref{generalQu}). Actually on can simply extract each of
them by means of a contour integral as written e.g. in \cite{torusIII}. 
 
On the other hand if we find a set of accessory parameters and of the
real parameter $\kappa$ such that the monodromies along all cycles and
around all sigularities are $SU(1,1)$ then expression
(\ref{expvarphi}) provides a single valued solution of Liouville which
we know to be unique. The above reasononig shows that we can replace
the problem of solving the Liouville equation to the one of finding a
set (which we know to be unique) of accessory parameters which make
all monodromies $SU(1,1)$.


In this section we shall write a minimal set the relations which
determine the $\beta$'s and the $\kappa$. All those parameters are
necessary to determine the solution. We shall first find a set of
relations which are sufficient to determine the $\beta$'s and do not
involve the $\kappa$. Then we give a relation which determines the
$\kappa$. We remark also that $\kappa$ intervenes always in the
combination $\kappa\bar\kappa$ and thus it counts only as one real
parameter.

We saw that the number of independent $\beta$ appearing in $Q_u$ are
$n+3g-3$. These correspond to $2n+6g-6$ real degrees of freedom. 
After choosing the elliptic monodromy at $u_K$ for $K=1$
diagonal, $q_1=D_1$ we have an additional real degree of freedom given
by $\kappa$ which describes the remnant $SL(2,C)$ transformation.

We have now to use such $2n+6g-5$ real degrees of freedom to make all
monodromies $SU(1,1)$. We know from the existence and uniqueness
theorem that this can be done and in a unique way.

Here we want to examine how this comes about. For clearness we start
from the case of genus $g=0$ (the sphere).

For $n=3$ there is no freedom of choice and from the explicit solution
in terms of hypergeometric functions (see e.g. \cite{MV}) we know that
a proper choice of the $\kappa$ makes $q_2$ $SU(1,1)$. Then using
$q_1 q_2 q_3=1$ we deduce that also that $q_3\in SU(1,1)$.
For $n=4$ we have one $\beta$ and we can exploit the $\kappa$ and one
real degree of freedom of $\beta$ to reduce $q_2$ to the form
\begin{equation}
q_2= 
\begin{pmatrix}
m_{11} & m_{12}\\
\bar m_{12} & m_{22}
\end{pmatrix}~.
\end{equation}
We have from the $SL(2,C)$ and the elliptic nature of the transformation 
\begin{equation}
m_{11}m_{22} = 1+m_{12} \bar m_{12}\geq 1,~~~~~~~~m_{11}+m_{22}=
-2\cos\alpha_2=~{\rm real},~~~~ |2\cos\alpha_2|\leq 2
\end{equation}
which give $m_{22}=\bar m_{11}$ i.e. $q_2\in SU(1,1)$. We can now use
the remaining real degree of freedom to have in $q_3$
$n_{11}=\rho_1 e^{i\phi},~n_{22}=\rho_2 e^{-i\phi}$ and from the
reality of the trace we derive $n_{22}=\bar n_{11}$. On imposing now
\begin{equation}
{\rm tr} D_1 q_2 q_3 = -2\cos \alpha_4 = {\rm real}
\end{equation}
we obtain for $q_3$ $n_{21}=\bar n_{12}$ i.e. $q_3 \in
SU(1,1)$. Finally due to $q_1q_2q_3q_4=1$ we have also $q_4 \in
SU(1,1)$. Increasing $n$ i.e. the number of sources by 1 we gain a
further $\beta$ i.e. two real degrees of freedom and we proceed as
above.

\bigskip

For $g>0$, $n>0$ we have
$n+2g$ cycles related by the algebraic relation \cite{FK}
\begin{equation}\label{cyclerelation}
q_1 \dots q_n a_1 b_1 a^{-1}_1b^{-1}_1\dots a_g b_g a^{-1}_gb^{-1}_g=1~.
\end{equation}
After fixing $q_1=D_1$ diagonal, the $n+3g-3$ $\beta$'s and $\kappa$ give us
$2n+6g-5$ real degrees of freedom to impose the $SU(1,1)$
nature to the remaining $q_2\dots q_n a_1 b_1\dots a_gb_g$. We spend
$2(n-1)$ of them to make all the elliptic $q_2\dots q_n$ $SU(1,1)$ and
we spend the $3(2g-1)$ remaining degrees of freedom to make
$a_1b_1\dots a_n$ (but not $b_n$) $SU(1,1)$. The $3$ degrees of
freedom to make each such element $SU(1,1)$ are spent as follows: two
degrees for  making $m_{21}=\bar m_{12}$ and one for obtaining
$|m_{11}|=|m_{22}|$. Then from
\begin{equation}\label{cycle relation}
m_{11}m_{22}=m_{12}m_{21}+1
\end{equation}
we obtain $m_{22}=\bar m_{11}$.
Using then 
eq.({\ref{cyclerelation}) we reach the equation 
\begin{equation}\label{reducedcyclerelation}
b_n a_n^{-1}b^{-1}_n \in SU(1,1)
\end{equation}
which being already $a_n$ in $SU(1,1)$, imposes three real constraints
on $b_n$ turning it from $SL(2C)$ into $SU(1,1)$.

A different way to proceed is the following: instead of fixing
$q_1$ diagonal, leave it undetermined. In this way we have in addition
to the $2(n+3g-3)$ degrees of freedom of the $\beta$'s the $3$ real
degrees of freedom of $SL(2,C)/SU(1,1)$ in total $2n+6g-3$ degrees of
freedom. Spend $2(n-1)$ of them to make the elliptic $q_2,\dots q_n$ 
$SU(1,1)$ and $3(2g-1)$ to make $a_1,b_1\dots a_n$ (but not $b_n$)
$SU(1,1)$. Use the $2$ left over parameters to make in $b_n$
$m_{21}=\bar m_{12}$ after which $b_n$ assumes the form
\begin{equation}
b_n= 
\begin{pmatrix}
m_{11}/\rho & m_{12}\\
\bar m_{12} & \bar m_{11}\rho
\end{pmatrix}
\end{equation}
where $\rho$ is a real parameter. 
The relation 
\begin{equation}
{\rm tr}(q_2\dots q_n a_1b_1\dots b_ga_g^{-1}b^{-1}_g)=-2\cos\alpha_1
\end{equation}
is a fourth order equation in $\rho$. From the existence and
uniqueness theorem we know that one solution is $\rho=1$ and such a
value makes $b_n$ and thus also $q_1\in SU(1,1)$, due to
eq.(\ref{cyclerelation}).

\bigskip

We come now to the writing of the relations which determine the
$\beta$'s without involving the parameter $\kappa$. The relations which
solve the monodromies at $(u_K,w_K)$ (dynamical singularities) 
are given by
\begin{equation}\label{qrelations}
m_{12}(q_K) = \bar m_{21}(q_K)
\end{equation}
because due to the elliptic nature of the monodromy we have
\begin{equation}
m_{11}(q_K)+m_{22}(q_K)= - 2\cos\alpha_K = {\rm
  real},~~~~|2\cos\alpha_K|\leq 2
\end{equation}
and
\begin{equation}
m_{11}(q_K)m_{22}(q_K)=m_{12}(q_K)\bar m_{12}(q_K)+1\geq 1.
\end{equation}
The relations assuring the monodromy along the cycles $a_l$ are
\begin{equation}\label{abrelations}
m_{12}(a_l) = \bar m_{21}(a_l)   
\end{equation}
\begin{equation}\label{abrelationsdiagonal}
 m_{11}(a_l)\bar m_{11}(a_l)=m_{22}(a_l)\bar m_{22}(a_l)
\end{equation}
and the same for the $b_l$ with $l=1\dots g$.

Below we denote by $M_{jk}$ the monodromy transformation of the two
independent solutions $f_1,f_2$ appearing in eq.(\ref{expvarphi})
along the various cycles.
Each matrix element $M_{jk}$ is an analytic functions of the
$p_j,\beta_s$ where $p_j=(u_j,w_j)$ with $j$ running over the
$n$ dynamical and the $2g+1$ kinematical singularities and $s$ runs
on $3g-3+n$ values. This is the
outcome of the solution of the auxiliary differential equation
given by the convergent Volterra series.

After starting canonical at
$(u_K,w_K)$ with $K=1$ we have $q_1=D_1$ and we spend one degree of freedom to
have
\begin{equation}\label{abrelationsdiagonal}
 M_{12}(q_2)= {\rm real}\times \bar M_{21}(q_2)
\end{equation}
which can be written as
\begin{equation}\label{q2offdiag}
M_{12}(q_2)M_{21}(q_2)=\bar M_{12}(q_2)\bar M_{21}(q_2)~.
\end{equation}
We spend now $2(n-2+2g-1)$ real degrees of freedom to impose
\begin{equation}\label{ratios}
\frac{M_{12}(q_2)}{\bar M_{21}(q_2)}=\frac{M_{12}(x)}{\bar M_{21}(x)}
\end{equation} 
where $x=q_3,\dots q_n,a_1,b_1\dots a_g$, while $2g-1$ are needed to have in the
$a_1,b_1\dots a_g$ 
\begin{equation}\label{diagonal}
|M_{11}|=|M_{22}|~.
\end{equation}
For satisfying (\ref{q2offdiag}), (\ref{ratios}) and (\ref{diagonal})
we need $2n+6g-6$ real parameters which are furnished by the
$n+3g-3$ complex $\beta$'s. Then $b_g$ becomes $SU(1,1)$ through the
relation (\ref{cyclerelation}). 
We notice that $\kappa$ does not intervene in the above relations and
it is determined by
\begin{equation}\label{kappaequation}
\frac{1}{\kappa^2\bar\kappa^2}\frac{M_{12}(q_2)}{\bar M_{21}(q_2)}=1~.
\end{equation}
The relations
(\ref{q2offdiag},\ref{ratios},\ref{diagonal}) are not pure analytic relations
as in all of them the complex conjugate of an analytic function
appears. In technical terms it means that the equations 
(\ref{q2offdiag},\ref{ratios},\ref{diagonal}) i.e. 
\begin{equation}\label{q2offdiag2}
M_{12}(q_2)M_{21}(q_2)=\bar M_{12}(q_2)\bar M_{21}(q_2)
\end{equation}
\begin{equation}\label{ratios2}
\frac{M_{12}(q_2)}{M_{12}(x)}=\frac{\bar M_{21}(q_2)}{\bar M_{21}(x)}
\end{equation} 
\begin{equation}\label{diagonal2}
\frac{M_{11}(a_l)}{M_{22}(a_l)}=\frac{\bar M_{22}(a_l)}{\bar M_{11}(a_l)}
~~~~l=1\dots g,~~~~~~~~
\frac{M_{11}(b_l)}{M_{22}(b_l)}=\frac{\bar M_{22}(b_l)}{\bar M_{11}(b_l)}
~~~~l=1\dots g-1
\end{equation}
define a real analytic
variety \cite{dangelo}. 
In order to deal with it, it is useful to promote the real
variables ${\rm Re}\beta_s,~{\rm Im}\beta_s,~{\rm Re} u_j,~{\rm Im}
u_j$ to complex variables. An equivalent procedure, which is formally
more handy, is the polarization process \cite{BM,dangelo}
which consists in considering
the variables $\beta_s,~\bar\beta_s,~u_j,~\bar u_j$ and promoting the
$\bar \beta_s$ and $\bar u_j$ to the independent complex variables
$\beta^c_s~,u^c_j$. Then the results relative to the original problem
are obtained for $u^c_j=\bar u_j,\beta_s^c=\bar\beta_s$.
Each equation of the type (\ref{ratios2}) 
gives rise
to two independent relations of the type
\begin{eqnarray}\label{complexpair}
&&A(\beta,u)=\bar B(\beta^c,u^c)\nonumber\\
&&B(\beta,u)=\bar A(\beta^c,u^c)~.
\end{eqnarray}
On the other hand relations  of the type 
(\ref{q2offdiag2},\ref{diagonal2}) are self-conjugate
in the sense that they give rise to the single equation
\begin{equation}
C(\beta,u)=\bar C(\beta^c,u^c)~.
\end{equation}

Finally we notice that two self-conjugate relations are equivalent to one
``complex'' relation e.g.
\begin{equation}
C(\beta,u)=\bar C(\beta^c,u^c)~~~~~~~~D(\beta,u)=\bar D(\beta^c,u^c)
\end{equation}
can be written as
\begin{equation}
F(\beta,u)=\bar G(\beta^c,u^c)~~~~~~~~G(\beta,u)=\bar F(\beta^c,u^c)
\end{equation}
with $F(\beta,u)=C(\beta,u)+iD(\beta,u)$,
$G(\beta,u)=C(\beta,u)-iD(\beta,u)$. In this way the eqs.(\ref{q2offdiag2},
\ref{ratios2},\ref{diagonal2}) can be rewritten as $n+3g-3$ pairs of 
complex relations of the type (\ref{complexpair}).

\section{The real-analyticity of the accessory parameters}\label{analyticity}

In proving Polyakov relation it is necessary to exploit the real
analyticity of the dependence of the accessory parameters $\beta$ and
of the parameter $\kappa$ on the moduli $u_K$, $u_l$. Actually due to
relation ({\ref{kappaequation}) it is sufficient to prove the
real-analyticity of the $\beta$'s.

On the sphere for any collection of parabolic singularities and of
finite order elliptic singularities it was proven by Kra \cite{kra}
that the accessory parameters are actually real-analytic functions of
$u_K$. Finite order elliptic singularities is the discrete set with
source strength $\eta = (1-1/n)/2$. For $n\rightarrow \infty$ they
accumulate to the parabolic limit. 

We are however interested in
the case in which the elliptic singularities are arbitrary.

However in this case we have no proof of real-analyticity everywhere
and thus our analysis will be of local nature.

A prerequisite in the proof of real-analyticity of the $\beta$'s
exploiting the monodromy conditions of section \ref{counting} is the
continuity of the $\beta$ on the moduli $u_K, u_l$.


In \cite{CMS2} it was proven using Green function technique that, as
expected, the functions $\phi, \partial_u \phi, \partial^2_u \phi$ are
uniformly bounded in any region of the $u$ plane, obtained by
excluding finite disks around the singularities, with bounds which
depends continuously on $u_j$.  Thus taking contour integrals of
eq.(\ref{generalQu}) at a finite distance from the singularities we
have that the $\beta$'s are bounded functions of the $u_K,u_l$ when
$u_K,u_l$ vary in a small polydisk. Such a result combined with
continuity of eqs.(\ref{q2offdiag2},\ref{ratios2},\ref{diagonal2}) and
the uniqueness of the solution implies that the $\beta$'s are
continuous functions of the $u_K,u_l$. Continuity is the basic
requirement to translate the equations of the previous section into the
local analysis of analytic varieties \cite{whitney}.

In the papers \cite{CMS1,CMS2} for the sphere topology it was proven
that the $\beta$'s are real-analytic function of the $u_K$ in an
everywhere dense open set in the space of the parameters $u_K$.

For clearness we illustrate the proof in the case of one accessory
parameter $\beta$, the extension to any number of accessory parameters
and moduli being straightforward.

In the following W stays for Weierstrass and WPT for Weierstrass
preparation theorem.  By Picard solution we understand the unique
values $\beta_R(u)$ $\beta_I(u)$ which solves the monodromy problem
(also in presence of parabolic singularities). The subscripts $R,I$
stay for the real and imaginary part.

We denote by $\Delta^{(i)}$ the set of relations assuring the $SU(1,1)$
nature of all monodromies.

Given a value $u_0$ we have $\Delta^{(i)}(\beta_R(u_0),\beta_I(u_0),u_0)=0$.  Let
$\Delta^{(1)}(\beta_R,\beta_I(u_0),u_0)$ be non identically zero in
$\beta_R$. Such $\Delta^{(1)}$ has to exist otherwise we violate the
uniqueness result. Then we can apply WPT to translate
$\Delta^{(1)}(\beta_R,\beta_I,u_R,u_I)=0$ into
\begin{equation}
P(\beta_R-\beta_R(u_0)|\beta_I,u_R,u_I)=0
\end{equation}
If $P$ is first order we have
\begin{equation}
\beta_R-\beta_R(u_0)+a_0(\beta_I,u_R,u_I)=0
\end{equation}
and $\beta_R$ is an analytic function of $\beta_I$ and $u_R,u_I$ in the
W-neighborhood ${\cal O}_0$ of $\beta_I(u_0),u_0$.
If $P$ is higher order
let us consider $P'$. 
If $P'(\beta_R(u_0)|\beta_I(u_0),u_0)=0$ but not identically zero in
a neighborhood of $u_0$, then we can solve for
$\beta_R$ for values of $u$ which lie as near as we want to $u_0$
and these form a open subset ${\cal O}_1$ of ${\cal O}_0$. 

If $P'(\beta_R(u)|\beta_I(u),u)\equiv 0$ in a neighborhood of $u_0$
then this constitutes a new equation which the Picard solution has to
satisfy and we proceed as above. Being the W-polynomial monic the
process ends in a finite number of steps and we have the result that
$\beta_R$ is an analytic function $\beta_R(\beta_I,u_R,u_I)$ of
$u_R,u_I$ and $\beta_I$, for points $u$ laying as near as we want to
$u_0$ and such set ${\cal O}_1$ is an open set.

We now consider $\Delta^{(2)}(\beta_R(\beta_I,u_R,u_I),\beta_I,u_R,u_I)
\equiv F(\beta_I,u_R,u_I)$ which is analytic in $\beta_I,u_R,u_I$.
Such  $F$ cannot be independent of $\beta_I$ otherwise 
$\beta_I$ would not be fixed by the monodromy conditions,
violating the uniqueness theorem. Then 
for any point $u\in{\cal O}_1$ we compute the W-polynomial
\begin{equation}
P(\beta_I-\beta_I(u_1)|u_R,u_I)=0
\end{equation}
and proceed as above. The result is that $\beta_R,\beta_I$ are
real-analytic functions of $u$ in an everywhere dense set.

Iterating, the above procedure works also when $u$ is any collection
$u_1,u_2\dots$ of parameters and we have any number of $\beta$. In
fact the existence and uniqueness result tell us that the
$\Delta^{(j)}$ fix completely the solutions.

\bigskip

In the case of a single accessory parameter like the torus with one
source or the four point problem on the sphere, a stronger result
can be obtained i.e. that the $\beta$ is a real-analytic function
everywhere except for a zero measure set in the $u$ plane \cite{torusIII}.

Through polarization \cite{BM,dangelo} i.e. promoting $\bar u$ and
$\bar \beta$ to 
new independent complex variables $u^c$ $\beta^c$, 
the single complex equation which
imposes the monodromy $A(\beta,u)=\bar B(\bar\beta,\bar u)$ is
promoted to a system of two equations
\begin{eqnarray}\label{polarizedsys}
&&A(\beta,u)=\bar B(\beta^c,u^c)\nonumber\\
&&B(\beta,u)=\bar A(\beta^c,u^c)~.
\end{eqnarray}
At the end  we shall be interested only 
in the self conjugate
solutions of the system (\ref{polarizedsys}) i.e. 
those which for $u^c=\bar u$ give $\beta^c=\bar\beta$.
We know such solution to exist and be unique. Applying WPT to
the two equations we have
\begin{equation}
P_1(\beta^c-\bar\beta(u_0)|\beta,u,u^c)=0
\end{equation}
\begin{equation}
P_2(\beta^c-\bar\beta(u_0)|\beta,u,u^c)=0~.
\end{equation}
A common solution of the two equations implies the vanishing of the resolvent
of the two polynomials
\begin{equation}
R(P_1,P_2)\equiv f(\beta,u,u^c)=0~.
\end{equation}
If $f(\beta,u_0,\bar u_0)$ vanish identically in $\beta$ the system
has solutions $\beta^c$ for any choice of $\beta$ near 
$\beta(u_0)$ but most important it can be easily proven
\cite{torusIII}
that we have infinite self-conjugate solutions for $u=u_0$, $u^c=\bar
u_0$ and $\beta$ near
$\beta(u_0)$. This violates Picard's uniqueness result.
Thus $f(\beta,u,\bar u)$ has to depend on $\beta$ and we can apply WPT
reducing it to the equation
\begin{equation}\label{resolventWpol}
P(\beta-\beta(u_0)|u,u^c)=0
\end{equation}
All the solutions of eq.(\ref{resolventWpol}), and in
particular the Picard solution, are analytic
in $u$ and $u^c$ i.e. real-analytic in $u$ except a zero measure
set as shown in the Appendix.

Thus for the case of a single accessory parameter, we have
real-analyticity of the Picard solution $\beta(u,\bar u)$
not only in an everywhere dense open set, but almost
everywhere in the space of the moduli.

\bigskip

In the general case of $N$ accessory parameters, as we have shown above,
real-analyticity holds in an everywhere dense open set, but we are not
aware of a proof of real-analyticity almost everywhere.

\section{Derivation of the Polyakov relation}\label{proof}

To derive Polyakov relation we shall go over to a finite
form for the action $S_u$ i.e. a form which does not contain
$\varepsilon\rightarrow 0$ limits as in eq.(\ref{SuactionGen}). 
This is achieved by decomposing the field $\varphi$ in a regular and singular
part similarly to what originally done in \cite{CMS1,CMS2}. 
Starting from
\begin{equation}
e^\varphi = \frac{2w_{12}\bar w_{12}}
{[\kappa^{-2}f_1\bar f_1-\kappa^{2}f_2\bar f_2]^2}
\end{equation}
where $f_1,f_2$ are solutions of
\begin{equation}
f''+ Q_uf=0
\end{equation}
we have near a singularity $u_K$ with $\zeta = u-u_K$
\begin{equation}
f''+\big(\frac{1-\lambda_K^2}{4\zeta^2}+\frac{\beta_K}{2\zeta}
+{\rm regular~terms}\big)f=0
\end{equation}
\begin{equation}
f_{1,2}=\zeta^{\frac{1\mp\lambda_K}{2}} y_{1,2}(\zeta)
\end{equation}
\begin{equation}
y_1''+\frac{1-\lambda_K}{\zeta}y_1'+
\big(\frac{\beta_K}{2\zeta}+{\rm regular~terms}\big)y_1=0
\end{equation}
\begin{equation}
y_1 = 1+ a\zeta+\dots;~~~~~~~~
a=-\frac{\beta_K}{2(1-\lambda_K)}=-\frac{\beta_K}{4\eta_K}~.
\end{equation}
Thus 
\begin{eqnarray}
&&e^\varphi=
{\rm const} (\zeta\bar\zeta)^{\lambda_K-1}
\big[(1+a\zeta+\dots)(1+\bar a\bar\zeta+\dots)\nonumber\\
&-&\kappa^4(\zeta\bar\zeta)^{\lambda_K}
(1+b\zeta+\dots)(1+\bar b\bar\zeta+\dots)\big]^{-2}
\end{eqnarray}
and then
\begin{equation}
\varphi = {\rm const}-(1-\lambda_K)\log \zeta\bar\zeta-2\big[a\zeta+\bar
a\bar\zeta+\dots -\kappa^4(\zeta\bar\zeta)^{\lambda_K} (1+b\zeta+\bar
b\bar\zeta+\dots)\big]
\end{equation}
where $1-\lambda_K=2\eta_K$.

Let $\Omega$ be a real field which is equal to $-2\eta_K \log(u-u_K)
(\bar u-\bar u_K)$ and to $-\frac{1}{2} \log(u-u_l) (\bar u-\bar u_l)$
in finite non overlapping disks around the singularities $u_K$, $u_l$
and equal to $-\frac{3}{2}\log u\bar u$ outside a disk of radius
$R_\Omega$ which includes all singularities. We shall call the union
of these regions $C$. Elsewhere $\Omega$ is defined as a smooth field
which connects smoothly with the field in the described
regions. Notice that $\Omega$ depends on the $u_K, u_l$.

In this way in the decomposition
\begin{equation}
\varphi = \varphi_M+\Omega
\end{equation}
$\varphi_M$ is finite and regular both at infinity and at the singularities.
Substituting such a decomposition in the action $S_u$ we obtain 
\begin{equation}\label{Sufinite}
S_u =\frac{1}{2\pi}\int(\frac{1}{2}\partial\varphi_M\wedge\bar
\partial\varphi_M - \varphi_M \partial\bar\partial \Omega-
\frac{1}{2} \Omega \partial\bar\partial \Omega+e^\varphi
du\wedge\bar d\bar u)\frac{i}{2}~.
\end{equation}
Varying $\varphi_M$ in (\ref{Sufinite}) we derive the equations of
motion for $\varphi_M$
\begin{equation}\label{varphiMeq}
\partial\bar\partial \varphi_M+\partial\bar\partial \Omega = e^\varphi
du\wedge d\bar u~.
\end{equation}
Due to the real-analytic dependence of the $\beta$'s and of $\kappa$
on the parameter $u_K$ 
the integrand in (\ref{Sufinite}) is
continuous in $u$ and $u_K$ and uniformly bounded by an integrable
function as $u_K$ varies is a small domain
and the derivative of the integrand w.r.t.
$u_K$ is continuous and bounded by an integrable function as $u_K$
varies is a small domain. Thus we can take the derivative under the
integral symbol
\begin{equation}\label{Sufiniteder}
\frac{\partial S_u}{\partial u_K}=\frac{i}{4\pi}
\int -\varphi_M \partial\bar\partial \Omega_K +
\Omega_K\partial\bar\partial  \varphi_{M} +\frac{1}{2}
\Omega_K\partial\bar\partial\Omega
-\frac{1}{2}\Omega\partial\bar\partial \Omega_K 
\end{equation}
where the subscript $K$ stays for 
$\displaystyle{\frac{\partial}{\partial
  u_K}}$. Notice that no $\displaystyle{\frac{\partial\varphi_M}{\partial u_K}}$ 
appears due to the equation of
motion (\ref{varphiMeq}). We
notice that in the disk around $u_K$ we have
\begin{equation}
\Omega_K = \frac{2\eta_K}{u-u_K},~~~~\bar\partial \Omega_K=0
\end{equation}
while in the remainder of $C$ we have $\Omega_K=0$ but  not
necessarily so in the complement of $C$.
The integral in eq.(\ref{Sufiniteder}) is finite and to perform the
integration by parts below it is useful to write it as the limit for 
$\varepsilon$ going to zero of the integral where a disk of radius
$\varepsilon$ around $u_K$ is excluded.
Integrating by parts we have
\begin{eqnarray}
&&\int\Omega_K\partial\bar\partial\varphi_M=
-\int
d(\Omega_K\partial\varphi_M)+\int\bar\partial\Omega_K\wedge\partial\varphi_M\\
&=&-\oint_{u_K}
\Omega_K\partial\varphi_M-\oint_{u_K} \varphi_M\bar\partial\Omega_K
+\int\varphi_M\partial\bar\partial\Omega_K=-\oint_{u_K}
\Omega_K\partial\varphi_M+\int\varphi_M\partial\bar\partial\Omega_K
\nonumber
\end{eqnarray}
where the last term cancels the first term in eq.(\ref{Sufiniteder}) and
\begin{equation}\label{contribution}
-\frac{i}{4\pi}\oint_{u_K}\Omega_K\partial\varphi_M=
-\frac{i}{4\pi}\oint_{u_K} \frac{2\eta_K}{u-u_K}(-2a)du=-\frac{\beta_K}{2}~.
\end{equation}
The minus sign is due the the fact that we are integrating on the
boundary of an inner domain.
Moreover we have
\begin{equation}\label{OmegaKterm}
\int\Omega_K\partial\bar\partial\Omega 
=
-\oint_{u_K}\Omega_K\partial\Omega+\int\bar\partial\Omega_K\wedge\partial\Omega
=\int\bar\partial\Omega_K\wedge\partial\Omega
\end{equation}
and
\begin{equation}\label{Omegaterm}
-\int\Omega\partial\bar\partial\Omega_K =-\oint_{u_K}
\Omega\bar\partial\Omega_K+\int\partial\Omega\wedge\bar\partial\Omega_K=
-\int\bar\partial\Omega_K\wedge\partial\Omega
\end{equation}
which cancels (\ref{OmegaKterm}). Summarizing
\begin{equation}\label{polyakovpfK}
\frac{\partial S_u}{\partial u_K}=-\frac{\beta_K}{2}.
\end{equation}
Parabolic singularities are treated in the same way with the same result.

We come now to the variation of the action under the variation of the
modulus $u_l$. The local uniformizing variable around $(u_l,0)$ is
$s$ with $s^2=u-u_l$. For $Q_s$ we have
\begin{equation}
Q_s= 2 B_l +O(s)
\end{equation}
and
\begin{equation}
B_l = \beta_l +\sum_K\frac{\epsilon_K w^2_K}{8(u_l-u_K)^2(u_l-u_1) 
\dots \{(u_l-u_l)\} \dots (u_l-u_{2g+1})}
\end{equation}
is the total accessory parameter at $u=u_l$.
The two independent solution of $f''+Q_s f=0$ around $s=0$ are given
by
\begin{equation}
f_1= 1+a_1s- B_l s^2+ a_3 s^3+\dots~~~,~~~~~f_2= b_1s+b_2s^2+ b_3 s^3+\dots
\end{equation}
For the $\varphi$ we have
\begin{eqnarray}
\varphi &=& -\frac{1}{2}\log(u-u_l)(\bar u-\bar u_l)-
2 \big[a_1s+\bar a_1\bar s - (B_l+\frac{a_1^2}{2})s^2-(\bar B_l+\frac{\bar
    a_1^2}{2}) \bar s^2
\nonumber\\
&-&\kappa^4 b_1 \bar b_1 s\bar s + O(s^3)]+{\rm const}~.
\end{eqnarray}
Then for the analogue of the integral (\ref{contribution}) we have
\begin{equation}\label{contributionmoduli}
-\frac{i}{4\pi}\oint^d_{u_l}\Omega_l\partial\varphi_M=
\frac{i}{4\pi}\oint_{0} \frac{1}{s^2}\bigg(a_1ds-(2B_l+a_1^2) s ds 
-\kappa^4 b_1\bar b_1 \bar s d s + O(s^2)ds\bigg) = - B_l -\frac{a_1^2}{2} 
\end{equation}
and thus
\begin{equation}\label{polyakovpfl}
\frac{\partial S_u}{\partial u_l}=-B_l -\frac{1}{8}(\partial_s \varphi_M)_{s=0}^2
~.
\end{equation}

The factor two of difference between eq.(\ref{polyakovpfK}) and
eq.(\ref{polyakovpfl}) in the coefficient of $B_l$ 
is due to the fact that the boundary of a disk
around $u_l$ in the $u$ cut-plane is a double turn.

\section{Conclusions}\label{conclusions}

Polyakov relation plays an important role in several aspects of
Liouville theory like the semiclassical limit of conformal blocks
\cite{ZZ,piatek,hadasz1,hadasz2}, the generalized monodromy problem
\cite{LLNZ,NRS} and the hamiltonian formulation of $2+1$ dimensional
gravity in presence of matter \cite{CMS1,CMS2,CMS3}.

I this paper we have extended Polyakov relation to all hyperelliptic
surfaces with an arbitrary number of sources. For higher genus we have
a relation between the accessory parameters and the change of the
action induced not only by the change in the position of the sources
but also by the change of the moduli.

After imposing the fuchsian conditions the number of independent
accessory parameters is $n+3g-3$ being $n$ the number of the sources and
$g$ the genus of the surface, and they are determined by imposing the
monodromy condition around the dynamical singularities and along the
fundamental cycles.

In the proof, as it happens already in the simple case of the sphere
it is necessary to exploit the real-analyticity of the accessory
parameters as functions of the singularities $u_K$ and $u_l$ which
represent the position of the sources and the moduli of the surface.

For the case of parabolic and finite order elliptic singularity we
know that such real-analyticity property is true everywhere
\cite{kra}.  For a collection of parabolic and arbitrary elliptic
singularities we proved that real-analyticity holds in an everywhere
dense open set in the space of the parameters $u_K$, $u_l$. For the
case of the torus with a single source and for the four point case on
the sphere we have the stronger result \cite{torusIII} that
real-analyticity holds everywhere except for a zero-measure set in the
space of the parameters.

Polyakov relation is then simply proven after decomposing the field in
a background component, which takes into account the singularities and
the behavior at infinity of the Liouville field, and a regular
part. With such a decomposition the change of the action reduces to the
computation of a single contour integral.

\section*{Appendix} 

In this appendix we derive the analytic properties of the solutions of
the equation given by the W-polynomial
\begin{equation}\label{weierappendix}
P(\beta-\beta(u_0)|u,u^c)=(\beta-\beta(u_0))^m
+a_{m-1}(u,u^c)(\beta-\beta(u_0))^{m-1}+
\dots +a_0(u,u^c)=0
\end{equation}
which appears in section \ref{analyticity}. 
The $a_j(u,u^c)$ are analytic functions
of $u$ and $u^c$ with $a_j(u_0,\bar u_0)=0$.

We start by computing the the resultant $R(P,P')=f(u,u^c)$ i.e. the
discriminant of $P$. We have two cases:

\bigskip

1. $f(u,\bar u)\not\equiv 0$ in the W-neighborhood of
$u_0,\bar u_0$. Then $f(u,u^c)$ can vanish only on a
``thin'' set \cite{whitney}. Such a set has zero 4-dimensional
Lebesgue measure \cite{GR} and the set where $f(u,\bar u)=0$ has zero
2-dimensional Lebesgue measure, as shown at the end of this appendix.

Thus except for such zero measure set we can apply the analytic
implicit function theorem to have $\beta(u,u^c)$ analytic function of
$u,u^c$ i.e. $\beta(u,\bar u)$ real analytic function of $u$.

\bigskip

2. $f(u,\bar u)$ is identically zero in the W-neighborhood of
$u_0,\bar u_0$. Then by a theorem on polarization \cite{BM,dangelo} we
have that $f(u,u^c)$ is identically zero.

In this case we proceed by computing
the reduced Gram determinants $D_n$ of the power-vectors of the roots
\cite{whitney}
\begin{equation}
D_n=
\begin{vmatrix}
s_0&s_1&\cdots&s_{n-1}\\
s_1&s_2&\cdots&s_n\\
\cdots&\cdots&\cdots&\cdots\\
s_{n-1}&s_n&\cdots&s_{2n-2}
\end{vmatrix}
\end{equation} 
where
\begin{equation}
s_i = \xi_1^i+\xi_2^i+\cdots+\xi_m^i
\end{equation}
$\xi_k$ being the $m$ roots of $P$. 
Being $D_n$ a symmetric polynomial of the
roots it is a polynomial in the coefficients $a_k(u,u^c)$ and as such an
analytic function of $u,u^c$. Notice that $D_m=R(P,P')$ 
\cite{whitney}.

In the present case
\begin{equation}
D_m(u, u^c)\equiv 0
\end{equation}
and we compute $D_{m-1}$. If it is not identically zero it means that the
maximum number of distinct roots is $m-1$ and the set where they are
$m-1$ is open and given by subtracting from the initial open set the
zeros of $D_{m-1}$ which is a thin set and as such of zero measure.
In the region where the maximum number of distinct roots is reached
all the solutions of (\ref{weierappendix}) (local sheet) are analytic
\cite{whitney}, and in particular the Picard solution is analytic.

Suppose now that 
\begin{equation}
D_m=D_{m-1}\equiv 0~.
\end{equation}
Then we compute $D_{m-2}$ an proceed as above. 

The procedure ends due to the fact that $D_1\equiv m$. It corresponds
to the situation where we have only one $m$-times degenerate solution
i.e.
\begin{equation}
P(\beta-\beta(u_0);u,u^c) = (\beta-\beta(u,u^c))^m =0
\end{equation}
from which we have
$\beta(u,u^c)-\beta(u_0)=-
\frac{1}{m}a_{m-1}(u,u^c) $ which is analytic in $u, u^c$ and thus
$\beta(u,\bar u)$ real-analytic in $u$.

Thus the accessory parameter $\beta$ is an analytic function of
$u,u^c$ everywhere except for a thin set. The thin sets in $u,u^c$ have
zero 4-dimensional Lebesgue measure \cite{GR}. However we are
interested in the 2-dimensional measure in the $u$ for $u^c=\bar u$,
i.e. given a function $f(u,\bar u)$ analytic in both arguments, we are
interested in the measure of the points where it vanishes.

In is simpler to go over to the ``real'' variables $x=(u+\bar u)/2,
~~y=-i(u-\bar u)/2$ and write $f(u,\bar u)=
f_r(x,y)$ which is also analytic in $x$ and $y$. 
Given any point $(x_0,y_0)$ it is always possible
\cite{BM} to perform a real linear invertible change of
variables as to make the WPT applicable at that point.
Then we can write
\begin{equation}\label{weierpol}
f_r(x,y) = U(x,y)((x-x_0)^k+ a_{k-1}(y)(x-x_0)^{k-1}+\dots a_0(y))
\equiv U(x,y)P(x-x_0|y)
\end{equation}
with $a_n(y_0)=0$ and $U$ a unit. The polynomial in (\ref{weierpol})
for each $y$ can vanish only at a finite number of points (real $x$). Then
denoting with $\Xi$ the function which equals $1$ where its argument
vanishes and zero otherwise we have
\begin{equation}
\mu = \int dy\int dx ~\Xi[P(x-x_0|y)]= \int dy ~0= 0~.
\end{equation}
We can represent the region of the modulus $u$ as the union of a
denumerable set of open domains. We have a zero-measure set of possible
non real-analyticity points in each domain and the union of such
infinite zero measure set has zero measure.

We conclude that $\beta$ is a real-analytic function of $u$ except for
a set of zero 2-dimensional Lebesgue measure in the $u$ plane.

\vfill


\end{document}